\documentclass[a4paper,showpacs,showkeys,nofootinbib]{revtex4}
\usepackage{latexsym}
\usepackage{amsmath, amsthm, amssymb, bbm}

\theoremstyle{plain}
\newtheorem{proposition}{Proposition}[section]

\theoremstyle{definition}

\newtheorem{example}[proposition]{Example}

\theoremstyle{remark}

\begin{document}

\newcommand{\ud}{\mathrm{d}}
\newcommand{\Tr}{\mathrm{Tr}}
\newcommand{\del}{\partial}
\newcommand{\order}{\mathcal{O}}
\newcommand{\V}[1]{\mathbf{#1}}
\newcommand{\R}{\mathbbm{R}}
\newcommand{\betrag}[1]{\left| #1 \right|}

\title{Noncommutative Electrodynamics with covariant coordinates}
\author{Jochen Zahn}
\affiliation{ II.~Institut f\"ur Theoretische Physik, Universit\"at Hamburg, Luruper Chausse 149, 22761 Hamburg, Germany}
\email{jochen.zahn@desy.de}
\date{August 10, 2004}


\begin{abstract}
We study Noncommutative Electrodynamics using the concept of covariant coordinates. We propose a scheme for interpreting the formalism and construct two basic examples, a constant field and a plane wave. Superposing these two, we find a modification of the dispersion relation. Our results differ from those obained via the Seiberg-Witten map.
\end{abstract}
\preprint{DESY 04-091}

\maketitle

\section{Introduction}

Motivated by Gedanken experiments on limitations of the possible localization of experiments~\cite{DFR} and string theory in a constant background $B$-field~\cite{SW}, there recently has been a lot of interest in noncommutative field theories, especially noncommutative electrodynamics. This theory is usually studied using the Seiberg-Witten map~\cite{SW}. This is a mapping from commutative gauge fields $\tilde{A}_{\mu}$ to noncommutative fields $A_{\mu}$, where $\tilde{A}_{\mu}$ transforms as the usual electrodynamic vector potential: $\delta \tilde{A}_{\mu} = \del_{\mu} \alpha$. The noncommutative fields are expressed as a power series in the commutator
\begin{equation}
\label{eq:comm_rel}
  \theta^{\mu \nu} = - i [q^{\mu}, q^{\nu}]
\end{equation}
of the coordinates. One can then write the Langrangean of NCED in terms of the commutative field strength $\tilde{F}_{\mu \nu} = \del_{[\mu} \tilde{A}_{\nu]}$:
\begin{equation*}
  L = - \frac{1}{4} \tilde{F}_{\mu \nu} \tilde{F}^{\mu \nu} + \frac{1}{8} \theta^{\alpha \beta} \tilde{F}_{\alpha \beta} \tilde{F}_{\mu \nu} \tilde{F}^{\mu \nu} - \frac{1}{2} \theta^{\alpha \beta} \tilde{F}_{\mu \alpha} \tilde{F}_{\nu \beta} \tilde{F}^{\mu \nu} + \order(\theta^2).
\end{equation*}
It leads to a nonlinear equation of motion for $\tilde{F}$. This results in a modified dispersion relation for an electromagnetic wave in a constant magnetic background field $\V{B}$, as has been shown in~\cite{Guralnik,Cai} for the case of space-space noncommutativity. In~\cite{Abe} this analysis was extended to space-time noncommutativity. For each direction of propagation, two plane wave solution with different polarizations were found. They fulfilled (in natural units) the following dispersion relation\footnote{This is their formula (4.19) with $a=1$. The crucial term $\pm 1/4 \sqrt{A}$ then becomes $\pm 1/4 \sqrt{(e_T B_T)^2 - (\V{e_T} \cdot \V{B_T})^2} = \pm 1/4 \betrag{ \hat{\kappa} \cdot (\V{e} \times \V{B})}$.}:
\begin{equation}
\label{eq:abe}
  \omega = k \left( 1- \V{m_T} \cdot \V{B_T} +  \frac{3}{4} \hat{\kappa} \cdot (\V{e} \times \V{B} ) \pm \frac{1}{4} \hat{\kappa} \cdot (\V{e} \times \V{B} ) \right) + \order(\theta^2).
\end{equation}
Here $\V{m}$ and $\V{e}$ denote the ``magnetic'' (space-space) and ``electric'' (space-time) parts of the commutator [see~(\ref{eq:theta})]. $\hat{\kappa}$ is the unit vector in the direction of propagation. The subscript $T$ denotes the component transversal to $\hat{\kappa}$. The two possibilities correspond to the two different polarizations.

But the Seiberg-Witten formalism has some drawbacks: Usually $\tilde{F}_{\mu \nu}(x)$, as a gauge invariant quantity, is interpreted as the field strength at $x$. This is similar to the interpretation of the Weyl symbol $\phi_W(x) := \int \ud^4k e^{-ikx} \Tr \phi(q) e^{ikq}$ as the value of the field $\phi$ at $x$, as is often, at least implicitely, done in scalar noncommutative field theory. But, as has already been pointed out in~\cite{DFR}, an evaluation functional should be positive. This requirement is not met here. Moreover, electrodynamics via the Seiberg-Witten map, if restricted to finite order in $\theta$, is a local theory. This is in contrast to the fact that noncommutative spaces, and thus also field theories on them, are inherently nonlocal. In fact, this was precisely the motivation for their introduction in~\cite{DFR}. Furthermore, already in~\cite{SW} doubts about the general validity of the perturbative expansion in $\theta$ have been raised.

For these reasons, we are studying NCED with another formalism, the covariant coordinates introduced in~\cite{Wess}. In the next section we review this formalism, comment on its interpretation, and give two simple examples, a constant field configuration and a plane wave. In the third section we try to superpose these solutions and find a modification of the dispersion relation similar to~(\ref{eq:abe}), but qualitatively different in the sense that there is no dependence on the polarization. In the last section we summarize and give a short outlook.

\section{Noncommutative Electrodynamics}

In noncommutative Electrodynamics, the field strength is given by
\begin{equation*}
  F_{\mu \nu} = \del_{\mu} A_{\nu} -  \del_{\nu} A_{\mu} - i [A_{\mu}, A_{\nu}].
\end{equation*}
The field strength and the vector potential $A_{\mu}$ lie in the algebra generated by the coordinates $q^{\mu}$ subject to the commutation relation~(\ref{eq:comm_rel}). A particular realization would be a suitable algebra of functions on $\R^4$, with the Moyal product as multiplication, but for the moment we prefer to work in the abstract setting.

The field strength is gauge covariant, i.e. it transforms as
\begin{equation*}
  F_{\mu \nu} \mapsto \Lambda  F_{\mu \nu} \Lambda^{-1},
\end{equation*}
where $\Lambda$ is a unitary element of the algebra. Gauge invariant quantities (observables), are then obtained using the trace in the algebra:
\begin{equation*}
  \Tr g(X) F^{\mu \nu}.
\end{equation*}
Here $X^{\mu} = q^{\mu} + \theta^{\mu \nu} A_{\nu}$ are the covariant coordinates~\cite{Wess} and $g$ is a suitable test function. We construct $g(X)$ in analogy to the Weyl-Wigner-Moyal calculus. Thus let $g$ be a Schwartz function on $\R^4$. Then define
\begin{equation*}
  g(X) := \int \ud^4k \ \check{g}(k) e^{ik_{\mu} X^{\mu}},
\end{equation*}
where $\check{g}$ is the inverse Fourier transform of $g$. In order to have a well defined expression, it is crucial that $X^{\mu}$, and thus $A_{\mu}$, is self-adjoint. Now the problem arises, that the map $g \mapsto g(X)$ does not preserve positivity (this is already the case for $A=0$). In order to have a positive evaluation functional, we propose the following procedure: Choose a Schwartz function $f_0$ centered around $x=0$ and set $f_x(y) = f_0(y-x)$. Then $g_x(X) := f_x(X)^* f_x(X)$ is positive by construction and the map
\begin{equation*}
  F^{\mu \nu} \mapsto  \frac{\Tr g_x(X) F^{\mu \nu}}{\Tr g_x(X)}
\end{equation*}
is positive and normalized. It should be interpreted as the evaluation of $F^{\mu \nu}$ at $x$. $f_0$ encodes the localization properties of the detector used. It would of course be desirable to choose it such that the resulting uncertainty is minimal, but this is a difficult problem for general $A_{\mu}$. Thus we will work with general $f_0$ in the following. The conclusions we draw from concrete examples in the remainder of this paper do not depend on a particular choice of $f_0$.

The action is defined as
\begin{equation*}
  S := - \frac{1}{4} \Tr F_{\mu \nu} F^{\mu \nu}.
\end{equation*}
This yields the equation of motion
\begin{equation}
\label{eq:eom}
  \del_{\mu} F^{\mu \nu} - i [A_{\mu}, F^{\mu \nu}] = 0.
\end{equation}

In the following we construct two solutions of this equation, one corresponding to a constant field and one to a plane wave, and evaluate the corresponding field strength in covariant coordinates. These are, to our knowledge, the first applications of covariant coordinates in concrete examples, except for the solitonic solutions constructed in~\cite{bak}.

\begin{example}
\label{ex:background}
Setting
\begin{equation}
\label{eq:A}
  A_{\mu} := c_{\mu \nu} q^{\nu}, \qquad c_{\mu \nu} \in \R
\end{equation}
we obtain the field strength
\begin{equation*}
  F_{\mu \nu} = c_{\nu \mu} - c_{\mu \nu} + (c \theta c^T)_{\mu \nu}.
\end{equation*}
The covariant coordinates are
\begin{equation*}
  X_1^{\mu} = q^{\mu} + (\theta c)^{\mu}_{\nu} q^{\nu} = (\mathbbm{1} + \theta c)^{\mu}_{\nu} q^{\nu} 
\end{equation*}
but in fact we do not need them in order to compute the measured field strength:
\begin{equation*}
  \frac{ \Tr (g_x(X_1) F^{\mu \nu})}{ \Tr g_x(X_1)} = F^{\mu \nu}.
\end{equation*}
\end{example}

\begin{example}
\label{ex:wave}
A plane wave is given by the vector potential
\begin{equation}
\label{eq:plane_wave}
  a_{\mu} = b_{\mu} ( e^{- i k q} + e^{ i k q}), \qquad b_{\mu} \in \R.
\end{equation}
The resulting field strength is
\begin{equation*}
  f_{\mu \nu} = - i ( k_{\mu} b_{\nu} -  k_{\nu} b_{\mu} )  ( e^{- i k q} - e^{i k q} ).
\end{equation*}
In Lorenz gauge $(k_{\mu} b^{\mu}=0)$, the equation of motion is then solved for $k^2=0$. The corresponding covariant coordinates are
\begin{equation*}
  X_2^{\mu} = q^{\mu} + \theta^{\mu \nu} b_{\nu} ( e^{-ikq} +  e^{ikq} ).
\end{equation*} 
To evaluate the field strength we would have to compute (assuming $f_0$ to be real) 
\begin{equation}
\label{eq:wave_field_strength}
  \frac{\Tr (g_x(X_2) f^{\mu \nu})}{\Tr g_x(X_2)} = - i  k^{[ \mu} b^{\nu ]} \frac{ \int \ud^4p \ud^4p' \ \check{f}_0(p) \check{f}_0(p') e^{i(p-p')x} \Tr ( e^{-ipX_2} e^{ip'X_2} ( e^{- i k q} -  e^{i k q}) )}{\int \ud^4p \ud^4p' \ \check{f}_0(p) \check{f}_0(p') e^{i(p-p')x} \Tr ( e^{-ipX_2} e^{ip'X_2} )}.
\end{equation}
This is a rather complicated expression. In fact we are not going to compute it, since we are mainly interested in the frequency content, not the corresponding amplitudes (which depend on the choice of $f_0$). To evaluate the traces, we expand $e^{-ipX_2}$, finding
\begin{equation*}
   e^{- i pq - i p \theta b (e^{- ikq}+e^{ikq}) } = e^{-ipq} e^{- i p \theta b  \left( P(p \theta k) (e^{- i k q}+e^{i k q}) - i Q(p \theta k) (e^{-i k q}-e^{i k q}) \right) },
\end{equation*}
with $P(x)=\frac{\sin x}{x}$ and $Q(x)=\frac{\cos x -1}{x}$. We express $e^{ip'X_2}$ analogously and write the exponentials of plane waves as power series, e.g.
\begin{equation*}
  e^{-i p \theta b P(p \theta k) \left( e^{-ikq} + e^{ikq} \right)} = \sum_n \frac{ \left( - i p \theta b P(p \theta k) \right)^n}{n!} \left( e^{-ikq} + e^{ikq} \right)^n.
\end{equation*}
Due to $\Tr e^{ikq} = (2\pi)^4 \delta(k)$, the traces will give sums of $\delta$ functions that set $p-p'$ to integer multiples of $k$. Since $\theta$ is antisymmetric, one then has $p \theta k = p' \theta k$ and the ``wavy'' parts of $ e^{-ipX_2} e^{ip'X_2}$ can be combined to
\begin{equation*}
  e^{i (p'-p) \theta b  \left( P(p \theta k) (e^{- i k q}+e^{i k q}) - i Q(p \theta k) (e^{-i k q}-e^{i k q}) \right) }.
\end{equation*}
Expanding this in a power series again, we get
\begin{equation*}
  \frac{\Tr (g_x(X_2) f^{\mu \nu})}{\Tr g_x(X_2)} = \\ - i k^{[ \mu} b^{\nu ]} \frac{\int \ud^4 p \ \check{f}_0(p) \check{f}_0(p-k) ( e^{\frac{i}{2} p \theta k} e^{-i kx} - e^{- \frac{i}{2} p \theta k} e^{i kx} ) }  {\int \ud^4 p \ \check{f}_0(p) \check{f}_0(p)} + \order(k \theta b).
\end{equation*}
At zeroth order in $k \theta b$ we thus find a plane wave with wave vector $k$, as expected. At $n$th order in $k \theta b$ we get a sum of plane waves with wave vectors $(n'+1)k, n' \leq n$, i.e., higher harmonics appear. In real experiments $k \theta b$ is of course a very small quantity. For the peak field strength of the proposed TESLA XFEL beam~\cite{tdr}, it can be estimated to be of order $10^{18}~\text{m}^{-2}~\lambda_{NC}^2$, where $\lambda_{NC}$ is the scale of $\theta$, which is supposed to be smaller than the scale reached by present-day accelerators.

The appearance of higher harmonics looks like a testable prediction, but there are some conceptual difficulties. First of all, we do not know wether what we produce in a laboratory is really of the form~(\ref{eq:plane_wave}). Adding higher harmonics to~(\ref{eq:plane_wave}) would still yield a solution of the wave equation. If the corresponding field strength is then evaluated in covariant coordinates as in~(\ref{eq:wave_field_strength}), the higher harmonics might at least partially cancel. Furthermore, as already mentioned, the amplitudes depend on $f_0$. But the exact correspondence between the detector and $f_0$ is not known. Thus the theory does not bear much predictive power concerning the higher harmonics. But obviously it is possible to determine the wave vector $k$ of the plane wave~(\ref{eq:plane_wave}) by local measurements of the field strength. We will exploit this in the next section.
\end{example}

\section{Electromagnetic waves in a constant background field}

Since the equation of motion~(\ref{eq:eom}) is nonlinear, the superposition principle does not hold any more. Nevertheless, it is possible to superpose the constant background field from example~\ref{ex:background} with a plane wave of the form~(\ref{eq:plane_wave}). But we will see that the wave vector is then in general no longer lightlike.

We define the complete vector potential as
\begin{equation*}
  A^c_{\mu} = A_{\mu} + a_{\mu},
\end{equation*}
where $A_{\mu}$ is given by~(\ref{eq:A}) and $a_{\mu}$ is of the form~(\ref{eq:plane_wave}). The field strength is then
\begin{equation*}
  F^c_{\mu \nu} = F_{\mu \nu} + \del_{\mu} a_{\nu} - \del_{\nu} a_{\mu} - i [A_{\mu}, a_{\nu} ] - i [a_{\mu}, A_{\nu} ] = F_{\mu \nu} + \del'_{\mu} a_{\nu} - \del'_{\nu} a_{\mu}.
\end{equation*}
Here we used
\begin{equation*}
  \del'_{\mu} g := \del_{\mu} g - i [A_{\mu}, g].
\end{equation*}
Inserting this in the equation of motion~(\ref{eq:eom}), we get
\begin{equation*}
  \del_{\mu} \left( \del'^{\mu} a^{\nu} - \del'^{\nu} a^{\mu} \right) - i \left[ A_{\mu},  \left( \del'^{\mu} a^{\nu} - \del'^{\nu} a^{\mu} \right) \right] = \del'_{\mu} \left( \del'^{\mu} a^{\nu} - \del'^{\nu} a^{\mu} \right) = 0.
\end{equation*}
Thus in the pseudo Lorenz gauge $\del'_{\mu} a^{\mu} = 0$, we get as equation of motion
\begin{equation*}
  \Box' a^{\nu} = 0.
\end{equation*}
In order to solve it, we seek the coordinates $q'$ dual to the derivatives $\del'$, i.e.
\begin{equation*}
  \del'_{\mu} q'^{\nu} = \delta_{\mu}^{\nu}.
\end{equation*}
Up to an additive constant these are
\begin{equation*}
  q'^{\mu} = { ( \mathbbm{1} - \theta c^T )^{-1} }^{\mu}_{\nu} q^{\nu}.
\end{equation*}
The equation of motion is thus solved by
\begin{equation*}
  a_{\mu} = b_{\mu} ( e^{-ikq'} +  e^{ikq'} ) .
\end{equation*}
with $k \cdot b = 0$ (pseudo Lorenz gauge) and $k^2 = 0$. The complete field strength is now
\begin{equation*}
  F_c^{\mu \nu} = F^{\mu \nu} - i (k^{\mu} b^{\nu} - b^{\mu} k^{\nu}) ( e^{-ikq'} - e^{ikq'} ).
\end{equation*}
Evaluating this in covariant coordinates, we see that the first term gives again the constant field strength $F^{\mu \nu}$. In the second term, we are once more only interested in the frequency content. The computation is completely analogous to the one in the preceding section, we simply have to replace $pq$ by $p(\mathbbm{1}+\theta c)q$ and $kq$ by $k (\mathbbm{1}-\theta c^T)^{-1} q$. Thus $p-p'$ is set to nonzero integer multiples of
\begin{equation*}
  k' = k (\mathbbm{1}-\theta c^T)^{-1} (\mathbbm{1}+\theta c)^{-1} = k (\mathbbm{1} - \theta c^T + \theta c - \theta c \theta c^T )^{-1} = k (\mathbbm{1}- \theta F)^{-1},
\end{equation*}
which is then the wave vector that is actually measured. First of all, we see that, contrary to~\cite{Abe}, it is independent of the polarization. The matrix $\theta F$ is expressed using the electric and magnetic parts of $\theta$ and~$F$:
\begin{equation}
\label{eq:theta}
  \theta^{\mu \nu} = \begin{pmatrix} 0   & -e_1 & -e_2 & -e_3 \\
                                     e_1 &  0   & m_3  & -m_2 \\
                                     e_2 & -m_3 & 0    & m_1  \\
                                     e_3 & m_2  & -m_1 & 0    \end{pmatrix}, \quad
  F_{\mu \nu} = \begin{pmatrix} 0   & E_1  & E_2  & E_3  \\
                                -E_1 &  0   & -B_3 & B_2  \\
                                -E_2 & B_3  & 0    & -B_1 \\
                                -E_3 & -B_2 & B_1  & 0    \end{pmatrix}.
\end{equation}
Hence, we find
\begin{align*}
  k_0 & = k'_0 ( 1 - \V{e} \cdot \V{E} ) - \V{k'} \cdot ( \V{m} \times \V{E} ) \\
  \V{k} & = \V{k'} ( 1 - \V{m} \cdot \V{B} ) + k'_0 ( \V{e} \times \V{B} ) - ( \V{k'} \cdot \V{e} ) \V{E} + ( \V{k'} \cdot \V{B} ) \V{m}.
\end{align*}
Assuming $\V{E}=0$ and $k_0' >0$, this leads to
\begin{align*}
  k_0 & = k'_0 \\
  \betrag{\V{k}}^2 & =  \betrag{\V{k'}}^2 \left( 1- 2 \V{{m}_T} \cdot \V{B_T} + 2 \hat{\kappa} \cdot (\V{e} \times \V{B}) \right) + \order((\theta B)^2),
\end{align*}
where $\hat{\kappa}$ is the unit vector in the direction $\V{k'}$. Using $k^2=0$ we find the modified dispersion relation
\begin{equation}
\label{eq:disp_rel}
  \omega' = k' \left( 1 - \V{{m}_T} \cdot \V{B_T} + \hat{\kappa} \cdot (\V{e} \times \V{B}) \right) + \order((\theta B)^2).
\end{equation}
This coincides with one of the two possibilities in~(\ref{eq:abe}). Considering only space-space noncommutativity (setting $\V{e}=0$), we are in agreement with the results obtained in~\cite{Guralnik,Cai}.

In order to discuss possible experimental test of the modified dispersion relation~(\ref{eq:disp_rel}), we estimate orders of magnitude. We consider a magnetic field of $1~\mathrm{T}$ and assume the noncommutativity scale to be close to the scale of present-day accelerators: $\lambda_{NC} = 10^{-20}~\mathrm{m}$. We find $\theta B \approx 10^{-24}$. Since this is a tiny number, it seems to be necessary to consider astronomical experiments, where small effects have enough time to sum up. One might for example use galactical magnetic fields. In the milky way these are of the order of $10^{-9}~\mathrm{T}$, so the correction would be of order $10^{-33}$. Multiplying by the diameter of the milky way, $10^{5}~\mathrm{ly}$, we find a shift in the arrival time of the order $10^{-20}~\mathrm{s}$. This seems to be far too small to be detectable. There is also the conceptual problem of finding a reference signal. Similar considerations can be found in~\cite{Guralnik}.

\section{Conclusion}

We proposed a scheme for interpreting noncommutative electrodynamics in terms of covariant coordinates and provided two simple examples: a constant field and a plane wave. We then tried to superpose these solutions and found a propagation speed different from the speed of light. Comparison with the results obtained in~\cite{Abe} via the Seiberg-Witten map shows a qualitative disagreement: In our setting the speed is not polarization dependent. 

Further research on this subject could proceed along the following lines: First of all, one should try to understand the discrepancy with the results obtained via the Seiberg-Witten map. One should also try to find other solutions and possibly measurable consequences. More ambitious goals would be a quantization of the theory and the treatment of nonabelian gauge groups, especially those different from $U(n)$.

\begin{bf} Added note: \end{bf} The authors of \cite{Abe} have checked their calculation and are now in agreement with the results presented here. I thank Y. Abe for this communication.

\begin{acknowledgements}
I would like to thank Klaus Fredenhagen for valuable comments and criticism. Financial support from the Graduiertenkolleg ``Zuk\"unftige Entwicklungen in der Teilchenphysik'' is gratefully acknowledged.
\end{acknowledgements}

\end{document}